\documentclass[3p,times,procedia]{elsarticle}
\usepackage{nupha_ecrc}
\usepackage{graphicx}

\volume{00}

\firstpage{1}

\journalname{Nuclear Physics A}

\runauth{I. Vitev}

\jid{nupha}

\jnltitlelogo{Nuclear Physics A}

\usepackage{amssymb}

\usepackage[figuresright]{rotating}

\begin{document}

\begin{frontmatter}



\dochead{XXVIIth International Conference on Ultrarelativistic Nucleus-Nucleus Collisions\\ (Quark Matter 2018)}

\title{
Quarkonium tomography of heavy ion collisions at the LHC
}


\author{I. Vitev}

\address{Los Alamos National Laboratory, Theoretical Division,
Mail Stop B283, Los Aalmos, NM 87545, USA}

\begin{abstract}
Quarkonium production in high-energy hadronic collisions provides a fundamental test of QCD. Its modification in a nuclear medium is a sensitive probe of the space-time temperature profile and transport properties of the QGP, yielding constraints complementary to the ones obtained from the quenching of light hadrons and jets, and open heavy flavor. In these proceedings, we report new results for the suppression of high transverse momentum charmonium [$J/\psi,\, \psi(2S)$] and bottomonium [$\Upsilon(1S),\, \Upsilon(2S),\, \Upsilon(3S)$] states in Pb+Pb collisions at the Large Hadron Collider. Our theoretical formalism combines the collisional dissociation of quarkonia, as they propagate in the quark-gluon plasma, with the thermal wavefunction effects due to the screening of the $Q\bar{Q}$ attractive potential in the medium. We find that a good description of the relative suppression of the ground and higher excited quarkonium states, transverse momentum and centrality distributions is achieved, when comparison to measurements at a center-of-mass energy of 2.76 TeV is performed. Theoretical predictions for the highest Pb+Pb center-of-mass energy of 5.02 TeV at the LHC, where new experimental results are being finalized are presented. Preliminary calculations for smaller systems, such as Xe+Xe are also shown. Last but not least, the potential of jet substructure to shed light in the mechanisms of heavy flavor production is discussed. 
\end{abstract}


\end{frontmatter}


\section{Introduction}
\label{intro}

Dissociation of the $J/\psi$ and the $\Upsilon$ meson families
in the quark-gluon plasma (QGP) created in heavy
ion collisions (HIC)  can help us characterize its properties. In particular, quarkonia are
sensitive to the space-time temperature profile and  transport coefficients of the QGP,
as recently summarized in~\cite{Andronic:2015wma}. 
Experimentally, an important observable that carries such information is the nuclear
modification factor of the yields of quarkonia in nucleus-nucleus ($AA$) collisions, when
compared to their yields 
in nucleon-nucleon ($NN$) collisions scaled with the number of binary interactions
\begin{equation}
  R_{AA}= \frac{1}{\langle N_{\rm coll.} \rangle }
  \frac{d \sigma_{AA}^{\rm Quarkonia}/dy dp_T}{d \sigma_{pp}^{\rm Quarkonia}/dy dp_T}    \, .
\label{raa}  
\end{equation}

The theory of quarkonium production is  non-relativistic quantum  chromodynamics 
(NRQCD)~\cite{Bodwin:1994jh}  and it is applicable at moderate transverse momenta.
It is expected that in  HICs the very short distance formation dynamics of a $Q\bar{Q}$
pair is not affected since $m_Q\gg T$, where $T$ is the typical QGP temperature.  
As the proto-quarkonium states propagate in matter, they interact with its quasiparticles.
The process of  collisional dissociation reduces the observed cross sections in the final 
state and the formalism was first developed for open heavy flavor, i.e. $D$-mesons and 
$B$-mesons~\cite{Adil:2006ra,Sharma:2009hn}.  In Ref.~\cite{Sharma:2012dy} this approach
was  extended to  quarkonia, though it failed to quantitatively reproduce the large $J/\psi$ 
suppression in central Pb+Pb reactions at the LHC and the larger attenuation of the 
excited $\Upsilon$ states.   One conceptual change in this framework is that now we solve for the
evolved quarkonium wavefunction and square the overlap with the thermal wavefunction to
get the survival probability~\cite{Aronson:2017ymv}.  This way, we also include the screening of the
color interaction between $Q$ and $\bar{Q}$ in a deconfined QGP~\cite{Matsui:1986dk}.

Production of quarkonia bears certain similarities with the production of open heavy flavor
and its modification in the QGP that we are just beginning to explore. We also present 
a new way to study the QCD dynamics of heavy quarks using heavy flavor tagged jet 
substructure~\cite{Li:2017wwc}.

\section{Theoretical formalism}
\label{formal}

\begin{figure}[!t]
\vspace*{.0in}
\includegraphics[width=2.8in,angle=0]{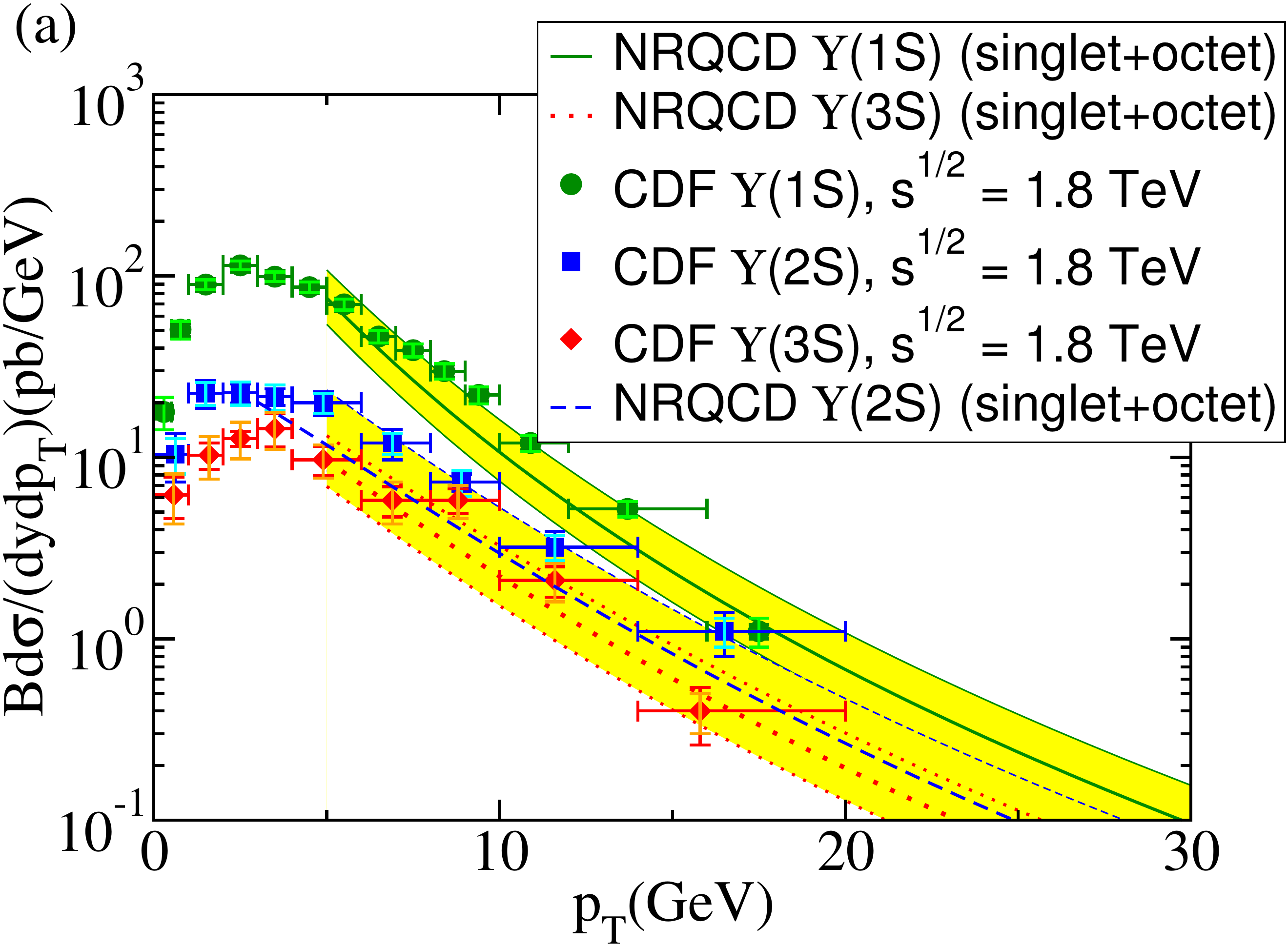}  \hspace*{.2cm}
\includegraphics[width=2.9in,height=1.9in]{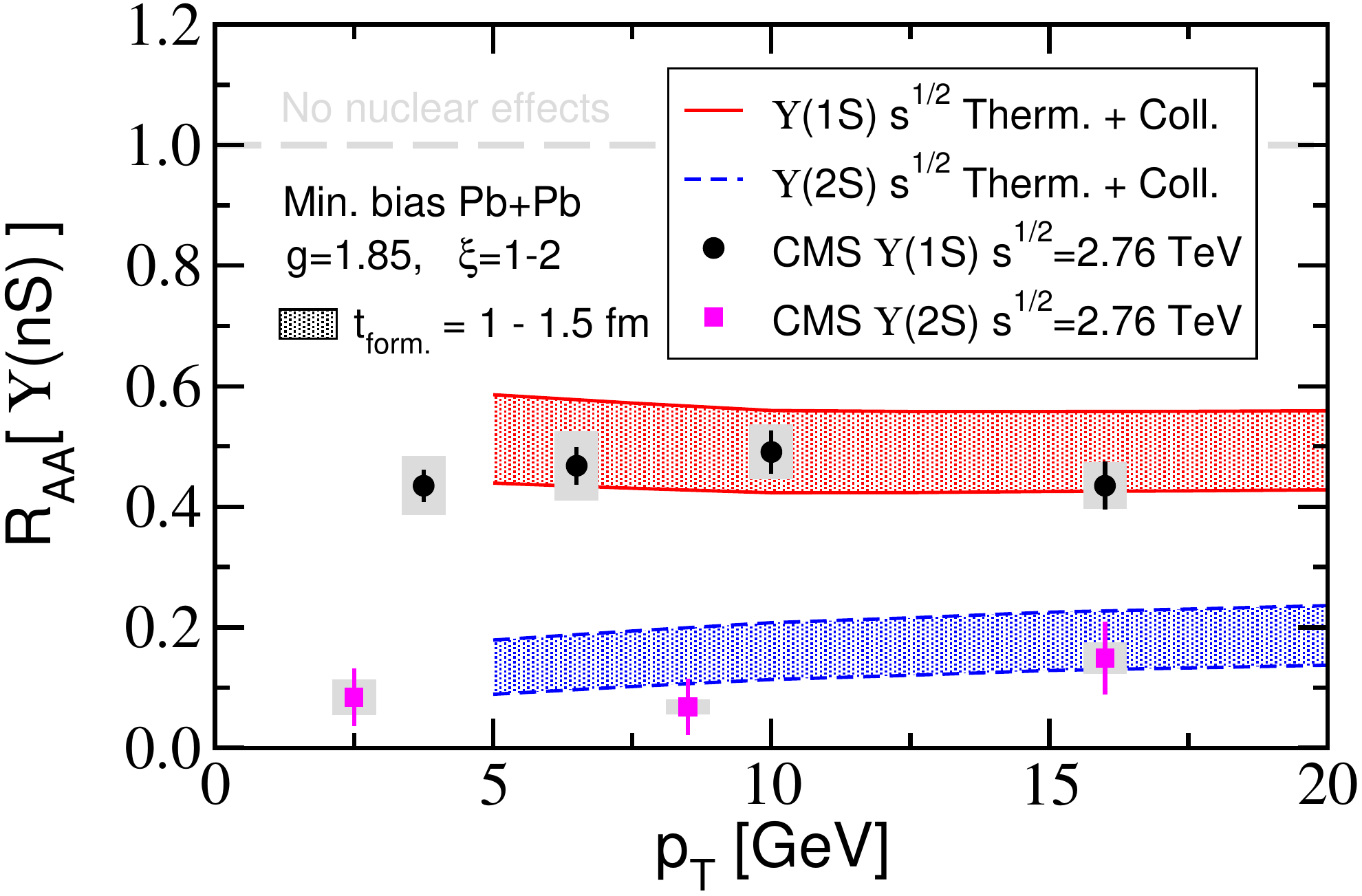} 
\caption{ Left panel:  $\Upsilon$ yields multiplied by
$B(\Upsilon\rightarrow\mu\mu) \simeq B(\Upsilon\rightarrow ee)\simeq 2.48\%$
for all three states. Theory comparison to data from the TeVatron at
$1.8$~TeV~\cite{Acosta:2001gv} is shown.  Right panel: Comparison of theoretical results for the
  $\Upsilon(nS)$ $R_{AA}$ in 2.76~TeV minimum bias Pb+Pb collisions versus $p_T$
to CMS experimental measurements~\cite{Khachatryan:2016xxp}.
}
\label{U276}
\end{figure}

In NRQCD, the $Q\bar{Q}$ pair is produced in a color-singlet or an octet state with a specific spin and orbital structure. 
It evolves into a quarkonium state with probabilities that are given by long distance matrix elements (LDMEs). For
color-octet states, this evolution process also involves the emission of soft
partons to form a net color-singlet object which we assume occurs on a
time scale which is shorter than ${\cal O}(1\, {\rm fm})$.  We build upon the NRQCD calculation of~\cite{Sharma:2012dy} 
and use the LDMEs extracted there to obtain good
description of the cross sections for bottomonia for $pp$ and
$p\bar{p}$ collisions for $p_T$ in the range of $5$ to $30$~GeV. For charmonia
we improve the $\chi_c$ fitting procedure by allowing the singlet matrix
element to be a free parameter. We also refit the $\psi (2S)$ by using
the LHC 7~TeV and CDF 1.8~TeV data. An example of the $\Upsilon (nS)$ cross sections we get in comparison to the
TeVatron data~\cite{Acosta:2001gv} is given in the left panel of Fig.~\ref{U276}. We note that precise $p_T$-dependent
feeddown from excited to lower-lying quarkonium states in our work is enabled by the NRQCD calculation.

As mentioned in the introduction, we take into account thermal screening and collisional dissociation of quarkonia. 
We determine the $J/\psi$ ands $\Upsilon$ families' properties by  solving the  the Schr\"{o}dinger equation at 
zero and finite temperature.  The wavefunctions $\psi_{i}(\Delta {\bf k}, x)$ of  proto-quarkonia are initialized with a 
width $\Lambda_0 \equiv \Lambda(T=0) $. 
This is a natural choice since in the absence of a medium it will evolve on the time-scale of
${\cal O}(1{\rm fm})$ or greater into the 
observed heavy meson. By propagating in the medium this initial wavefunction  accumulates transverse momentum 
broadening  $\chi \mu_D^2 \xi$.  The probability that this  $Q\bar{Q}$ configuration will transition into a
final-state heavy meson 
with thermal wavefunction $\psi_{f}(\Delta {\bf k}, x)$ with $\Lambda(T)$ is given by
\begin{eqnarray}
  P_{f\leftarrow i} (\chi\mu_D^2 \xi,T) & = & \left|  \frac{1}{2 (2\pi)^{3} }  \int d^{2}{\bf k} dx \,
\psi_{f}^* (\Delta {\bf k},x)\psi_{i}(\Delta {\bf k}, x) \right|^{2}  
\nonumber \\
&& \hspace*{-1.in}= \left| \frac{1}{2 (2\pi)^{3} }  \int dx \; {\rm Norm}_f {\rm Norm}_i \, \pi  
  \,  e^{-\frac{ m_{Q}^{2} }{ x(1-x)\Lambda(T)^{2} } }  e^{-\frac{ m_{Q}^{2} }{ x(1-x)\Lambda_0^{2} } }    \,   \frac{ 2 [ x(1-x)\Lambda(T)^{2}]
[\chi\mu_D^{2}\xi+x(1-x)\Lambda_0^{2}] }
{   [ x(1-x)\Lambda(T)^{2}] + [\chi\mu_D^{2}\xi+x(1-x)\Lambda_0^{2}]  }  \; \right|^2 \, . \;\; \quad 
\label{sprob}
\end{eqnarray}
In Eq.~(\ref{sprob}) ${\rm Norm}_i$  is the normalization of the initial state,
including the transverse momentum broadening 
from collisional interactions,  and  ${\rm Norm}_f$ 
is the normalization of the final state. We describe production and dissociation of quarkonium states by a set of coupled 
rate equations\cite{Sharma:2012dy,Aronson:2017ymv}
that describe their evolution in a QGP modeled by  EBE-VISHNU (2+1)-dimensional 
event-by-event viscous hydrodynamic package~\cite{Shen:2014vra}.

\section{Quarkonium phenomenology}
\label{pheno}

\begin{figure}[!t]
\includegraphics[width=3.in,height=2.in]{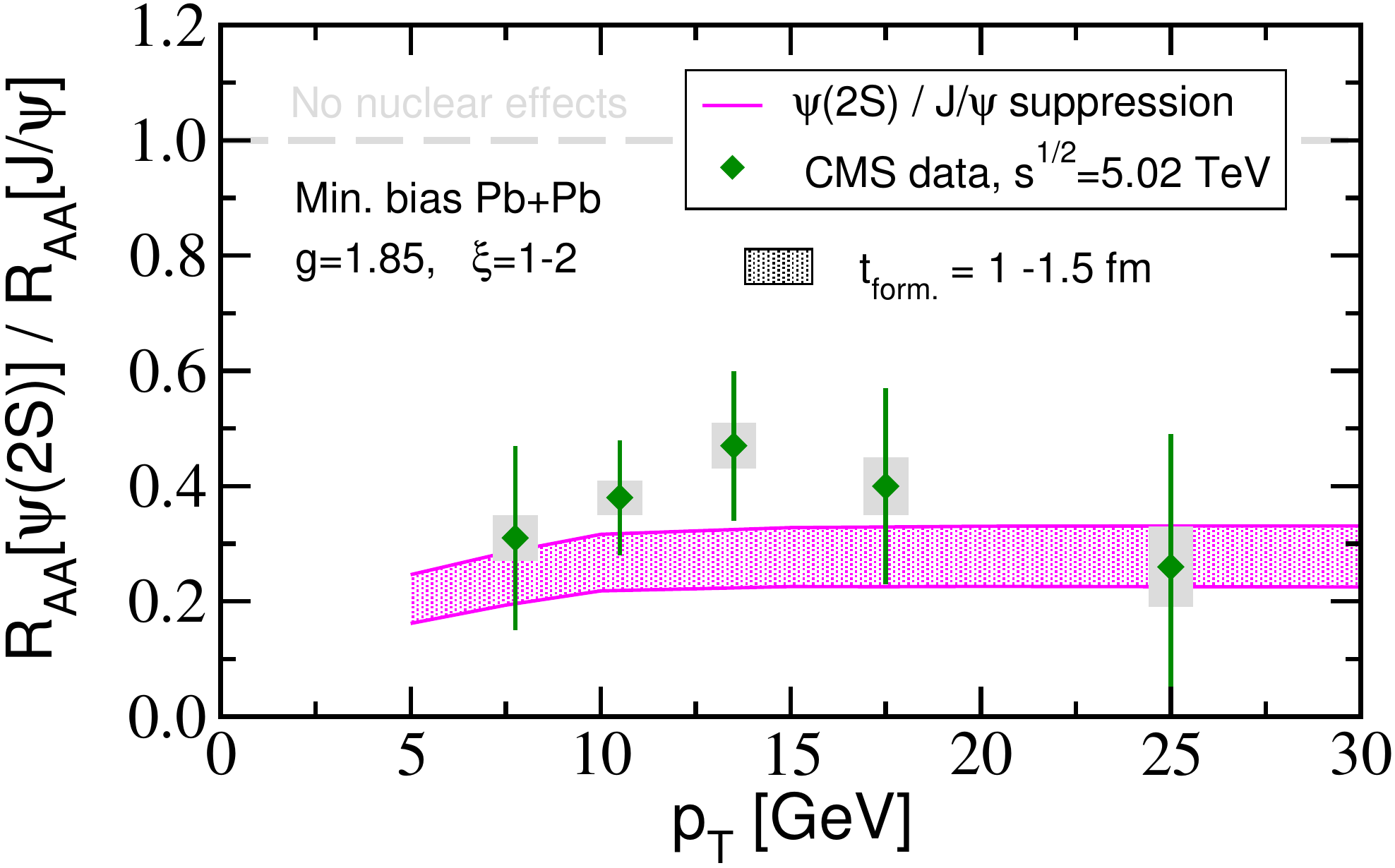}   \hspace*{.2cm}
\includegraphics[width=3.in,height=3.in]{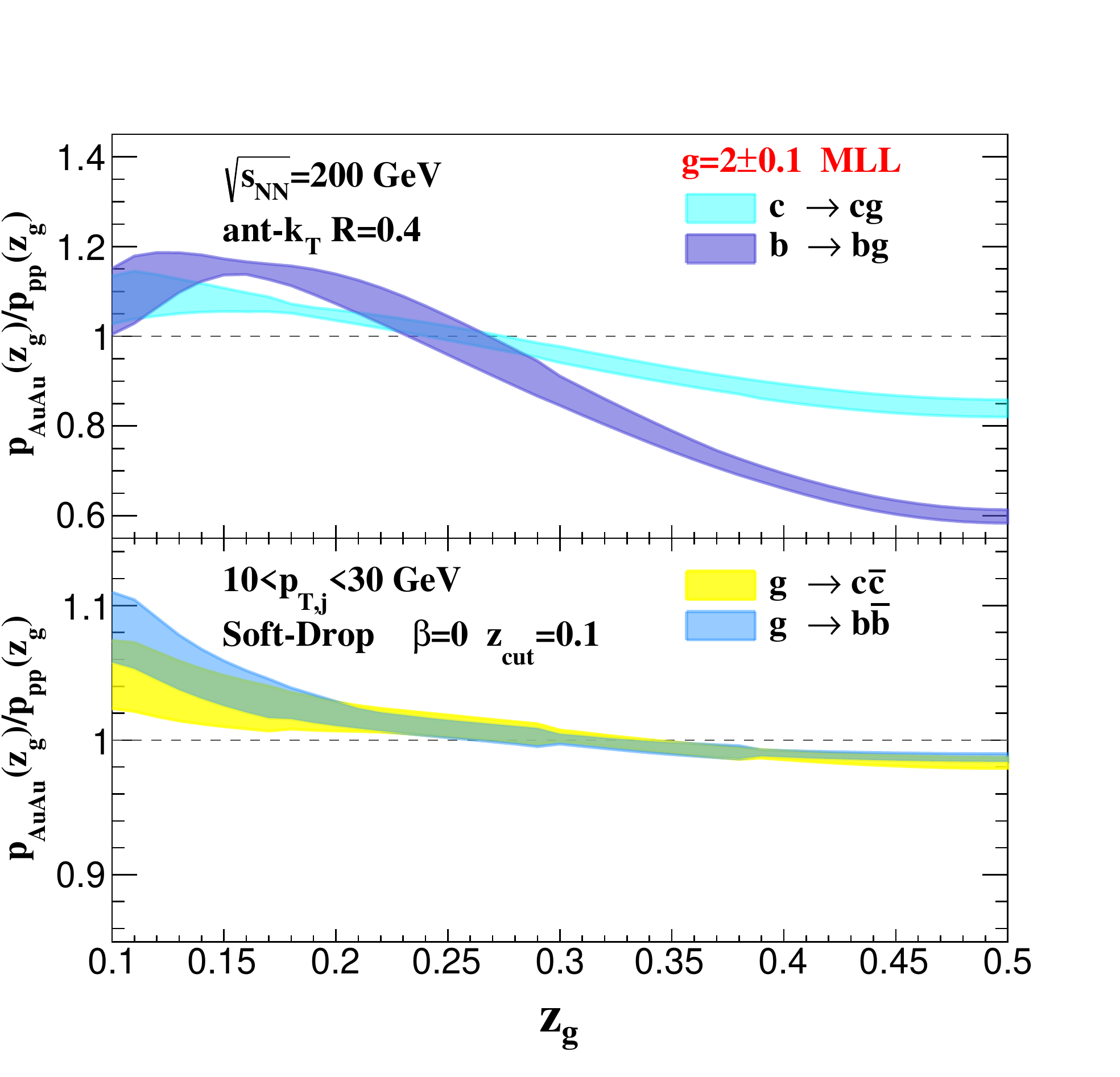} 
\caption{Left panel:  theoretical model predictions for the
  double $(\psi(2S)_{AA}/(\psi(2S)_{pp})/ (J/\psi_{AA}/(J/\psi_{pp})$ ratio in  minimum bias
  Pb+Pb collisions 
at  $\sqrt{S}=2.76$~TeV at the LHC. Data is from CMS~\cite{Sirunyan:2016znt}.  Right panel: The modifications of the splitting functions for heavy flavor tagged jet at $\sqrt{s_{\rm NN}}=200$~GeV Au+Au collisions. 
}
\label{J2toJ1}
\end{figure}

\
Theoretical results are obtained for all $J/\psi$ and $\Upsilon$ states in Pb+Pb collisions 
at center of mass energies 2.76~TeV and 5.02 TeV at the LHC~\cite{Aronson:2017ymv}. We find good 
agreement with  existing data as a function of transverse momentum and centrality and show several 
examples below. Recently, experimental measurements of the differential  suppression of the $\Upsilon(nS)$
family have appeared at high $p_T$~\cite{Khachatryan:2016xxp}. Theoretical calculations
for the $\Upsilon(1S)$ (red band) and $\Upsilon(2S)$ (blue band) in minimum bias
$\sqrt{s}=2.76$~TeV Pb+Pb reactions are shown in the right panel of Fig.~\ref{U276}. 
The experimental data is described well, including its magnitude and $p_T$ dependence.
We note that collisional dissociation mostly        
affects the ground  $\Upsilon$ state, while thermal wavefunction effects dominate
the attenuation pattern of the excited $\Upsilon$ states. The same is true for the $J/\psi$
states~\cite{Aronson:2017ymv}.
The bands reflect the combined uncertainty
of the interaction onset time $t_{\rm form.}$ and the collisional dissociation of
the quarkonium states. In the evaluation of the latter we keep the coupling between
the heavy quarks and the medium fixed at $g=1.85$~\cite{Sharma:2012dy} but vary the
broadening parameter $\xi$. The upper edge of the uncertainty band corresponds to
$t_{\rm form.} = 1.5$~fm, $\xi =1$. The CMS collaboration also
put an upper limit on the  $\Upsilon(3S)$ cross section in Pb+Pb reactions, corresponding to an
upper limit on its $R_{AA}$~\cite{Khachatryan:2016xxp}. Our calculated $\Upsilon(3S)$    
cross section is consistent with this limit.

At $\sqrt{s} =5.02$~TeV,  measurements of relative suppression
ratios of excited to ground charmonium states have
appeared to high $p_T$~\cite{Sirunyan:2016znt}. The data for $\psi(2S) / J/\psi $
is shown in the left panel of Fig.~\ref{J2toJ1}.  Our theoretical calculations
are compatible with the experimental data within the statistical and systematic error bars. The large
difference in the relative suppression of the ground and excited quarkonium states to the highest
measured $p_T$ is important because it provides essential constraints on an alternative 
picture of quarkonium suppression based on parton energy loss.

\section{Heavy jet substructure}
\label{substructure}

To better understand the mechanisms of heavy flavor production we now turn to the substructure of heavy flavor jets~\cite{Li:2017wwc}. 
Our work builds upon Ref.~\cite{Chien:2016led} but includes Sudakov resummation, which is especially important for the 
$g\rightarrow Q\bar{Q}$ channel.  In the kinematic domain where parton mass plays the most important role, we predict a unique inversion of the mass hierarchy of jet quenching effects, with the modification of the  momentum sharing distribution for prompt $b$-jets being the largest. 
Numerical results for the momentum sharing distribution ratios for heavy flavor tagged jets in  Au+Au to p+p collisions  at $\sqrt{s}=200$ GeV are presented in the right panel of Fig.~\ref{J2toJ1}.   For $c\to cg$, the  
$p(z_g)$ modification in the QGP is similar to the one for light jets, however, the $b\to bg$ channel  exhibits  much larger in-medium effects. 

\section{Conclusions}
\label{conclusions}

In summary, we presented theoretical results for the $p_T$-differential suppression of charmonia and bottomonia in Pb+Pb 
collisions at the LHC. Our calculations are based on NRQCD and capture the essential physics of thermal and collisional dissociation of $J/\psi$s and $\Upsilon$s   in the QGP~\cite{Aronson:2017ymv}. While an illustrative subset of results was presented here, detailed
predictions are available that will allow to test this model versus upcoming experimental measurements of quarkonium 
suppression at $\sqrt{s}=5.02$~TeV, as well as  smaller systems, such as Xe+Xe.  We also presented a novel method to constrain 
heavy flavor production and mass effects on parton shower formation in the QGP~\cite{Li:2017wwc}, which we expect will grow into
a broad program of heavy flavor jet substructure studies.









\end{document}